\documentclass[twocolumn,showpacs,floatfix,epsfig]{revtex4}
\usepackage[final]{graphicx}
\usepackage{epsfig}

\begin{document}

\title{Smallness of the imaginary part of $\eta$-nucleus scattering length}
\author{J.A. Niskanen}

\affiliation{
Department of Physical Sciences, PO Box 64, FIN-00014 University of
Helsinki, Finland }


\begin{abstract}
In two recent analyses of $\eta$ meson production a very
small imaginary part was obtained for the $\eta ^3$He
scattering length, which is in contradiction with most
theoretical predictions.
In this report a plausible explanation is
given to this unexpectedly weak absorption by suppression of
the two main inelasticity channels. It is stressed that the
real and imaginary parts of the elementary scattering length do not
necessarily always have the same isospin algebraic and spatial
properties in a nuclear environment and caution should be exercised
in their use in multiple scattering calculations and even
constructing simple optical potentials.
\end{abstract}
\pacs{25.80.-e; 21.45.+v; 25.40.Ve; 13.75.-n}

\maketitle

Production of the $\eta$ mesons close to threshold has obtained
considerable interest during the last decade or so. One reason
is that there is a possibility of bound mesonic nuclei, i.e.
nuclei with an $\eta$ meson bound with the strong interaction
\cite{Liu}. This expectation is
based on the knowledge that the $\eta N$ interaction is quite
attractive with a significantly sizable scattering length.
However, there is a long standing controversy about how heavy
nuclei can bind an $\eta$ meson.  Most works give $^4$He or
heavier as the lowest limit, but some claim even $^3$He to
be able to bind an $\eta$ meson \cite{Chiang,Wilkin,Wycech,Rakityansky,
Fix1,Haider1,Haider2}.

The scattering quantity standardly associated with binding
is the sign of the real part of the scattering length, $a_R$.
With the convention normal in meson physics
\begin{equation}
q \cot\delta = \frac{1}{a}+ \frac{r_0}{2}q^2\, ,
\label{efran}
\end{equation}
a positive $a_R$ indicates moderate attraction, while a negative
value means repulsion or a bound state.
However, as pointed out earlier by Haider and Liu, also
$|a_R| > a_I $ should be valid \cite{Haider2}. By unitarity,
the imaginary part $a_I$ is always positive. In Ref.
\cite{Sibirtsev} this game was carried even further to second order
in $r_e/a$ with the condition
$a^3(a^* - r_{e}^*) > 0$, which reduces to the former one,
if $r_e=0$. From these conditions (albeit with the bold
assumption that $|a|\gg |r_e|$) one can see that also the
imaginary part of the scattering length has an essential
role even for the very existence of bound states, not to say
anything about their width. In fact, a very strong correlation
between $a$ and the binding energy and the width was seen
in Ref. \cite{Sibirtsev2} giving constraints for the latter
{\it provided bound states do exist}. For this reason
a detailed study and
understanding of also the imaginary part of the $\eta$-nuclear
scattering length is relevant.

In Ref. \cite{Sibirtsev} a reanalysis of existing data on
the $\eta \, ^3$He system was presented. These data stem from the
reaction $pd \rightarrow \eta\, ^3{\rm He}$ and the extraction
of the scattering length was based on the standard low energy
expression of the final state interaction
\begin{equation}
|f|^2=\frac{|f_p|^2}{1{+}a_I q {+}|a|^2 q^2}\, ,
\end{equation}
where the original production amplitude $f_p$ is assumed to
be very short ranged and essentially momentum independent.
The global fit to available data gave the result
$a = \pm 4.3 \pm 0.3 + i\, (0.5 \pm 0.5)$ fm.
It should be stressed that this analysis cannot determine
the sign of the real part, which only appears in the second power.
Further, this result is fully consistent with a coupled channel
$K$-matrix analysis of Ref. \cite{Green} yielding
$a= 4.24\pm 0.29 + i\, 0.72\pm 0.81$ fm.

Somewhat surprisingly  the above analysis gives only a very small
imaginary part, consistent with zero. Naively one might have
expected the result to be of the order of three times the elementary
amplitude $a_{\eta N}$ at least for the imaginary part. The
imaginary part of the elementary scattering length in turn varies in
the range 0.2 -- 0.4 fm \cite{Wycech}, so that the naive expectation
would be about 1~fm. Indeed, the results computed by various models
and various elementary inputs vary between 0.5 fm and 6 fm.

There are two main sources for the inelasticity in the system.
Firstly there is the inherent input inelasticity in the
elementary amplitude $a_{\eta N} \approx 0.5 \pm 0.2
+ i\, (0.35 \pm 0.1)$ giving the above mentioned 'naive' expectation.
Secondly, with the nucleus there is also the possibility
of nuclear break-up. In the following I will address both and
argue that both are significantly suppressed. \\

\noindent {\em Pionic inelasticity}:\\
The elementary inelasticity is due to the process
$\eta N \rightarrow \pi N$. It seems that most (if not all)
calculations merely consider the complex scattering length
as one entity. However, from the origin of the inelasticity
it is clear that the real (elastic) part and imaginary
(inelastic) part have a different isospin structure.
Only the elastic part is of the trivial identity operator
form in isospin, while the latter is necessarily of the
form $\pi\cdot\tau$ and leads to the one-nucleon operator
$\sum_i \pi\cdot\tau_i$ operating on the nucleons.
To the extent that the pion field spatially
could be approximated by unity (i.e. the plane wave argument
can be neglected), the result is simply the total isospin
operator. Since the considered nuclei have the isospin one
half, for the $^3$He as well as for the triton this gives
the same value as for a single nucleon and there is {\em no
factor of three}. So from the pionic inelasticity one should
not have more than the single nucleon value $\approx 0.3$ fm.
However, this very approximate assumption is not needed as
seen below.

The above quoted value is obtained actually for a very compact
object, a single baryon. In the case of a coherent process in a
nucleus actually the form factor of the nucleus arises as a function
of the momentum transfer $q_{\pi} \approx 490$ MeV$/c$ (we neglect
the $\eta$ momentum), which is a small and common multiplier to the
isospin operator of each nucleon and consequently the same total
isospin operator is obtained as above. This coherent (nucleus
conserving) reaction suppression by the form factor by even two
orders of magnitude is not taken into account in simple optical
potentials, where, in spite of the very different kinematics $a_R$
and $a_I$ are treated on equal footing.

Actually the softness of the nucleus does not come fully into the
suppression, because, in principle, the nucleon form factor could
attenuate this somewhat. Namely one might write the physically
measured effective elementary $a_I$ in terms of the $\eta N$ form
factor and a ``pure´´ coupling as
\begin{equation}
a_I=a_I({\rm eff}) = F_{\eta N}(q_\pi)\, a_I({\rm pure})
\end{equation}
and write the nuclear form factor effect symbolically
\begin{equation}
a_I(\eta A) = F_({\eta A}){q_\pi}\, a_I({\rm pure}) = F_{\eta
A}(q_\pi)/F_{\eta N}(q_\pi)\, a_I({\rm eff}) \, .
\end{equation}
However, $F_{\eta N}$ is much harder than $F_{\eta N}$ and in
practice the inelasticity from this source must be negligibly small.
Using Gaussian distributions giving r.m.s. radii 0.6 fm and 1.9 fm
for the nucleon and $^3$He, respectively, a suppression factor of
0.04. Together with the $a_I$ of $\eta N$ scattering this result is
supported by the data on $\pi^- \, ^3{\rm He} \rightarrow \eta t$ of
Ref. \cite{Peng89} as can be seen from Fig. 6 of Sibirtsev {\it et
al.} \cite{Sibirtsev}.

The above exercise is to some extent a repetition of the fact, known
long ago in pion production and absorption in nuclei, that processes
involving large momentum transfers are small with a single nucleon.
On the other hand, quite clearly the charge exchange inelasticity
$\pi^\pm A\rightarrow \pi^0 A$ at threshold does not have this
massive suppression, since a large momentum transfer is not
necessary and the reaction and elastic channels are
kinematically nearly the same. \\

\noindent {\em Nuclear inelasticity}:\\
In general with large momentum transfers it is economical
to share it with typically two nucleons. In {\it e.g.} pion
absorption on nuclei the absorption on two nucleons is
by far dominant. We consider this now.

It is an experimental fact that the cross section for
$np \rightarrow np\eta$ is about 6.5 times as large as
that for $pp \rightarrow pp\eta$ \cite{etadiff}.
This is very suggestive, since if also in $\eta$ absorption on
bound nucleon pairs (with the same quantum numbers as in the
final states of the mentioned reactions)
\begin{equation}
\sigma_{np} = \frac 1 2 \, (\sigma_0 + \sigma_1 )
\approx 6.5\sigma_{pp} = 6.5\sigma_1 \, ,
\end{equation}
then it follows that absorption on an isospin one pair is
suppressed by an order of magnitude compared with isospin
zero. This is very similar to pion absorption also
on $^3$He, where by far the dominant process is on a quasideuteron
in the three-nucleon system rather than on a singlet pair.

It may well be that pseudoscalar mesons prefer absorbing
on triplet pairs when possible.
Further one may note that the isoscalar pair wave function
part is somewhat favoured in $^3$He \cite{Baru}. In view of
these facts we may safely neglect absorption on the $pp\,$
or in general on a singlet pair in attempts to understand
the smallness of $a_I$ and concentrate on the isosinglet
component alone.

At this stage one may invoke information from the reaction
$np \rightarrow d\eta$. For the absorption reaction one would
get
\begin{equation}
\sigma({\eta d\rightarrow np}) = \frac 4 3 \frac{p^2}{q^2}\,
\sigma(np \rightarrow d\eta )\, ,
\end{equation}
with $p$ the nucleon and $q$ the $\eta$ momentum.
This in turn can be used in the optical theorem
\begin{equation}
f(\theta = 0) = \frac {q}{4\pi}\, \sigma_{\rm tot}
\end{equation}
to give the contribution from this source to the imaginary part of
the forward amplitude, i.e. $a_I$ in the $s$ wave. The numerical
extrapolation to threshold from Ref. \cite{Calen1} would be about
0.01 fm, while the available lowest energy results of Ref.
\cite{Calen2} are rapidly varying (indicating a large scattering
length) with the zero energy limit being 0.030 -- 0.035 fm. In
quasifree absorption of pions on quasideuterons in $^3$He a factor
of 1.5 -- 2 was found a long time ago both theoretically
\cite{quasith} and experimentally \cite{quasiex}. This factor may
account for the denser wave function and the fact that in $^3$He
there are 1.5 isoscalar pairs. A similar factor is expected for
$\eta$ absorption and one would gain perhaps 0.05 - 0.07 fm from
this source to the imaginary part of the scattering length.

Also this means  that once the $\eta$ meson has been absorbed
on a dominant isoscalar pair, this pair cannot emit a pion
staying in an $S$ state. Namely
the $s$-wave nature of the elementary process tends to
preclude the
change of the spin and parity and consequently also the isospin.
This Pauli blocking effect (albeit in a small scale with only two
nucleons) has been hinted at in Refs. \cite{Wycech,Hanhart}

However, since the elementary process does not take place at the
centre of mass of the two active nucleons, the quasifree absorption
$\eta d \rightarrow NN\pi$ is in principle
possible with the final nucleons in a relative triplet $P$ state
and pion in the $p$ state relative to the two nucleons
(to conserve the parity, $d$ denotes here the quasideuteron pair).
The final $Pp$ state is not very
favourable as can be seen from the following estimate.
For an order of magnitude estimate one may use plane waves for
the pion and high energy nucleons
\begin{equation}
M \propto \int_0^\infty j_1(pr)\, j_1(qr/2)\, \psi_d(r)\, r^2\, dr \, ,
\end{equation}
where $\psi_d$ is the quasideuteron pair wave function for
which I use the parameterization of Ref. \cite{Baru} for the
CD Bonn potential \cite{bonn}. (Here $q$ is the pion momentum.)
This amplitude is squared and integrated over the phase space.
The result is a suppression by a factor of 0.54 as compared with
the pion and nucleons coming out in $s$ waves and by a
factor of 0.026 as compared with elastic zero energy
scattering.

Another possible nuclear reaction or decay channel  $\eta\, {^3}{\rm
He} \rightarrow pd$  was already studied in Ref. \cite{Sibirtsev}
and found to be able to contribute less than 0.01 fm. There remains
still three-nucleon absorption. However, in pion absorption this has
always been smaller than absorption nucleon on pairs
\cite{quasiex,Weber} and there is no reason to expect this mode to
be stronger here. However, this point remains for checks in $\eta$
absorption reactions, since, obviously, it cannot be studied by
inverse production reactions as the other channels considered
above.\\

\noindent {\em Summary} \\
In conclusion, I have considered some reasons why the originally
quite large pionic inelasticity of $\eta N$ scattering decouples in
the case of nuclei. The real and imaginary parts of the $\eta N$
scattering length actually have a different isospin dependence
and momentum dependence, which changes the ratio of the elastic
and reaction channels in a nuclear environment. This is not taken
into account in calculations predicting large $a_I$'s. Genuine
absorption on the nucleus does not contribute very much either.

With the suppression of both one-nucleon pionic
inelasticity and estimates of the absorption on nucleon pairs, it
seems quite understandable that $\eta ^3$He scattering should be
rather elastic as indicated in the analyses of Refs.
\cite{Sibirtsev,Green}. The largest inelastic contribution seems to
be the nuclear decay due to the quasifree absorption on an
isosinglet (quasideuteron) pair ${\eta d\rightarrow np}$, which
gives probably 0.05 -- 0.07 fm to the imaginary part of the
scattering length. Other channels appear negligible as compared
to this and there seems no need for $a_I$ larger than 0.1 fm. It
would be of interest to see if the prediction can be experimentally
confirmed or whether three-nucleon decay becomes comparable to
quasi-two-body absorption.

By generality of the arguments this conclusion is probably valid
also for $\eta$-meson interactions with other nuclei, notably
$^4$He, and might be generalized also to some other mesonic
interactions. In particular, the prospects for bound $\eta$-nuclear
states may be increased with a small $a_I$.

More generally the present discussion shows that even in simplistic
calculations starting from some elementary (complex) amplitude,
often approximated by the scattering length, it should not
be forgotten that the real and imaginary parts normally
have different spin-isospin structures and may in nuclei
lead to different results than naively expected.

\begin{acknowledgments}
This work was supported by the DAAD and Academy of Finland exchange
programme projects DB000379 (Germany) and 211592 (Finland).
 I also thank the Magnus
Ehrnrooth Foundation for partial support in this work and
IKP of Forschungszentrum J\"ulich for hospitality.
\end{acknowledgments}

\end{document}